\documentclass[pre,showpacs,twocolumn]{revtex4}
\usepackage{graphicx}
\newcommand{\rb}{\mbox{\boldmath $r$}}
\newcommand{\Ob}{\mbox{\boldmath $0$}}

\newcommand{\be}{\begin{equation}}
\newcommand{\ee}{\end{equation}}

\begin{document}

\title{On the static length of relaxation and the origin of dynamic
heterogeneity \\ in fragile glass-forming liquids}

\author{S.\ Davatolhagh}
\affiliation{Department of Physics, College of Sciences,
 Shiraz University, Shiraz 71454, Iran}
\date{\today}

\begin{abstract}
The most puzzling aspect of the glass transition observed
in laboratory is an apparent decoupling of dynamics from
structure. In this paper we recount the implication of various
theories of glass transition for the static correlation length
in an attempt to reconcile the dynamic and static
lengths  associate with  the glass problem. We argue
that a more recent characterization of the static relaxation
length based on the bond ordering scenario, as the typical length
over which the energy fluctuations are correlated, is more
consistent with, and indeed in perfect agreement with the typical
linear size of the dynamically heterogeneous domains observed in deeply
supercooled liquids. The correlated relaxation of bonds in
terms of energy is therefore identified as the physical origin of
the observed dynamic heterogeneity.

\end{abstract}


\maketitle
\section{INTRODUCTION}\label{sec:1}
In the glassy phenomenology, it is the super-Arrehnius slowing down of
the transport properties that is most striking \cite{Ang96,Deb96}.
We refer to the so-called fragile liquids \cite{Ang91}, which are distinguished by a highly
temperature-dependent effective energy barrier, $E_{\rm eff}(T)$,
in their thermally activated expression for structural or $\alpha$
relaxation time
\be\tau_{\alpha}(T) = \tau_{\infty}\exp (E_{\rm eff}(T)/k_BT),
\label{eq:1.1}\ee
where $\tau_{\infty}\sim 10^{-13}$s
is a high-$T$ relaxation time, and $k_B$ is the Boltzmann constant. The temperature
variation of $\tau_{\alpha}(T)$ for fragile supercooled liquids is described
over a wide range of temperatures by the empirical Vogel-Fulcher (VF) equation \cite{VF}
\be \tau_{\alpha}(T) = \tau_{\infty}\exp [A/(T-T_0)]. \label{eq:1.2} \ee
The apparent divergence temperature $T_0$ appearing in Eq.\ (\ref{eq:1.2}), is called the
Vogel-Fulcher temperature, and often found to be very close to
the Kauzmann temperature $T_K$ \cite{Kauz} where the configurational entropy of
the liquid appears to vanish, if it were to stay in equilibrium \cite{Ang97}.
Although $T_K$ is not a rigorous lower-bound, it is a fairly good
indication as to the lowest temperature a liquid can be supercooled \cite{Ell90}.
We note that Eq.\ (\ref{eq:1.2}) automatically predicts a phase
transition at $T_0$ \cite{KTZK96,Mez01}. It is the case that in finite dimensions,
and short-range interactions, a diverging time is normally accompanied
by a diverging length. In fact, the large activation energy
barrier $E_{\rm eff}(T_g) \approx 50\;k_BT_g$ observed for weakly bonded
fragile liquids at the laboratory glass temperature $T_g$,
is regarded as an indication for the cooperative nature of the
relaxation dynamics \cite{TK00}. Several equilibrium theories of the
structural glass transition, invoke an increasing static
correlation length that diverges at $T_K$ ($\approx T_0$) \cite{AG65,GD58,KTW89}.
As the dynamics is activated, even a small correlation length leads to macroscopic
values for $\tau_{\alpha}$ exceeding the observation time: thus, the
falling out of equilibrium of the liquid at the laboratory glass
temperature $T_g$ ($>T_0$), which is conveniently defined as that
temperature where $\tau_{\alpha}(T_g) \sim 10^3$s \cite{Ang91}.

The most puzzling aspect of this super-Arrehnius dynamic slow
down is that it is not seen to be accompanied by any obvious
change in the static structure of the liquid.
The static structure factor (i.e.\ Fourier transform of the pair correlation
function) obtained by X-ray and neutron scattering experiments reveals only marginal
changes in the local density profile of polymers
and molecular supercooled liquids \cite{Scat}.
In other words, an `extended' short-range order of less than a nanometer in size begins
to appear as the liquid is cooled toward $T_g$. This occurs while the $\alpha$ relaxation
time as obtained from viscosity and a.c.\ dielectric susceptibility measurements increases
by as much as 12 orders of magnitude. It is reconciling
the above two aspects of the structural glass transition, i.e.\ structure versus dynamics,
which remains a major challenge in the condensed matter physics\cite{And95}.
Clearly, there is no detectable growing length
associated with the density fluctuation. However, there are general arguments in favor of
the fourth-order potential energy density correlation function and associated thermal
susceptibility, the structural specific heat, as the ideal quantities for
studying the static properties in fragile \cite{Dav05} and model glass-forming liquids \cite{Fer05}.
Although in experiments the contribution of the potential energy is difficult to separate from
the kinetic contributions to the specific heat, it is accessible in computer simulations of
model glass-formers such as prototype Binary Mixture of Lennard-Jones (BMLJ)
particles \cite{Kob98,Sas99}.

From another perspective, and unlike the simple liquids treated
as homogeneous \cite{HM86}, deeply supercooled liquids are also distinguished by
dynamic heterogeneous domains \cite{Edi00,Sil99}, a few nanometers across,
with widely different relaxation times (varying by five
orders of magnitude). Recent multi-dimensional Nuclear
Magnetic Resonance (NMR) measurements are found to favor a heterogeneity size,
$\xi_{\rm het}=2$--3 nm, i.e.\ 5 or more atomic diameters \cite{Edi00,Tra98}.
Results obtained from the fluctuation theory \cite{LL80} using the Heat
Capacity Spectroscopy (HCS) data \cite{Don82} are similar---though some times
higher \cite{Don01}. As for their dependence on the temperature, more recent
experimental procedures have discovered a growing dynamic length accompanying
the glass formation in colloidal and molecular liquids \cite{Ber05}.
However, the origin of this dynamic heterogeneity
remains unclear to date. Do dynamic heterogeneous domains correspond
to any static correlation in the structure? The heterogeneity of time
suggests possible heterogeneity in the structure.
Thus, a knowledge of the typical size of the static and/or dynamic
lengths, their temperature dependencies and correlation
(if any) can go a long way in resolving the structural glass problem \cite{Edi00}.

In a previous work \cite{Dav05,DP01}, we proposed a Bond Ordering (BO) scenario for
the glass transition in which the cooperative
relaxation of bonds in terms of energy, uncorrelated with density ordering or
crystallization \cite{Tan98,Tan99}, has been discussed.  In this bond ordering picture,
the structural component of the specific heat arising from the potential energy
fluctuation, plays the central role as the thermal susceptibility associated with
the static length for cooperative relaxation \cite{Dav05}. This static relaxation length,
$\xi_{\rm BO}$, is defined as that length-scale over which energy fluctuations
are correlated. It may also be regarded as an operational definition for the average
linear size of Cooperatively Rearranging Regions (CRRs) discussed in the context of
Adam-Gibbs-DiMarzio theory of cooperative relaxation \cite{AG65,GD58}.
The scenario predicts a growing
(and possibly diverging) structural specific heat with the lowering temperature for
the fragile liquids, which has been in part corroborated by more recent Monte Carlo
(MC) simulations of a Lennard-Jones binary mixture \cite{Fer05}, by Fernandez et\,al.
They found that local potential energy fluctuations
are in fact correlated over distances much larger than the short range of
the inter-atomic interactions. This is made evident by studying finite-size effects
at normally inaccessible temperatures $T<T_{\rm MCT}$, where $T_{\rm MCT}$ is the
apparent mode coupling transition temperature \cite{GS92}. Temperatures as low as
$0.895\;T_{\rm MCT}$ are accessed and studied in equilibrium using a local swap MC dynamics
that is insensitive to the `cage-effect', and thus drastically reduces the equilibration time.

Our aim  in this paper is to make evident the correlation between the static relaxation
length, $\xi_{\rm BO}$, defined as the average length over which energy fluctuation
is correlated, and the typical linear size of the dynamically heterogeneous domains,
$\xi_{\rm het}$, which close to $T_g$ is five or more atomic diameters \cite{Edi00,Ber05}.
This is done in an attempt to clarify the physical origin of dynamic heterogeneity.

The rest of this paper is organized as follows:
In Section \ref{sec:2}, we briefly review the implication for the static correlation length,
of the theories of structural glass that invoke an underlying ideal glass transition.
Prominent examples are the Adam-Gibbs-DiMarzio entropy model, and discontinuous mean-field
spin glass models that have a random first-order transition in the mean-field formalism.
We also discuss the potential-energy-landscape view of dynamics in supercooled
liquids in section \ref{sec:2}. This paves the way in section \ref{sec:3}
for a phenomenological determination of the structural
contribution to the specific heat, $C_s$, and the corresponding static
length of relaxation, $\xi_{\rm BO}$, for the fragile supercooled liquids.
The thermal behavior of $\xi_{\rm BO}$ is derived from that of $C_s$ via an
energy version of the Fisher scaling law.
In section \ref{sec:4}, we discuss the implication of this novel approach with regard to
the physical origin of dynamic heterogeneity, and its apparent relation to the mean field
models of structural glass transition. A brief summary of the main results appears in section \ref{sec:5}.

\section{Implication of previous theoretical work for static correlation length}
\label{sec:2}
\subsection{Adam-Gibbs entropy model}
\label{sec:2.1}
By considering the liquid to be made up of identical clusters of size $z$, each capable
of undergoing independent rearrangement, Adam and Gibbs (AG) \cite{AG65} arrived at an
activated expression for $\tau_{\alpha}(T)$ where the effective energy barrier is given by
$E_{\rm eff}(T) = z^*(T)\Delta\mu$. $\Delta\mu$ is to be largely interpreted as the
`potential energy barrier' per atom against rearrangement in a cluster composed
of $z^*(T)$ atoms, and taken to be a constant. $z^*(T)$ is the minimum `critical' size of a CRR,
allowing a transition between two configurations. By definition, the configurational entropy
of this minimal CRR is given by $s_c^*\equiv k_B\ln 2$. It is easy to see that the ratio of
the total number of atoms $N$ to the size of minimal CRR $z^*$, should be equal to the ratio
of the total configurational entropy $S_c$ to the minimum entropy of a CRR $s_c^*$:
$N/z^*(T) = S_c(T)/s_c^*$ \cite{Sil99}. Thus, $z^*(T)$ is expressed in terms of the configurational entropy of the system
\be z^*(T) = Nk_B \ln 2 / S_c(T) \label{eq:2.11}. \ee Furthermore, by assuming a hyperbolic form
$\Delta C_p = C/T$ for the excess specific heat of the liquid over the equilibrium crystal \cite{Ang97},
an approximate expression for the configurational entropy is obtained by the thermodynamic
integration \be S_c(T) = \int_{T_K}^T dT \Delta C_p/T = C(T-T_K)/T. \ee
On substitution into Eq.\ (\ref{eq:2.11}), this results in VF equation being recovered.
We also note that $z^*(T)\propto T/(T-T_K)$.
Letting $z^*(T) \sim \xi_{\rm AG}^3$, a (minimal) static relaxation length is obtained
\be \xi_{\rm AG}(T) = \xi_{\infty}[1-(T_K/T)]^{-1/3},\label{eq:2.12}\ee
where $\xi_{\infty}$ is constant and should be regarded as the typical
inter-atomic spacing or an atomic diameter.

In fact the concept of the increasing size of cooperative regions with the lowering temperature,
dates back to considerations by Jenckel in 1939 \cite{Jen39}. However, the thermal variation
of cooperativity was formulated in terms of the atomic packing energy, while AG theory
emphasizes the role of configurational entropy. It should also be pointed out that the
central results of the Adam-Gibbs theory could be obtained without recourse to the concept
of configurational entropy \cite{Moh94} with only the assumption of a diverging $z^*(T)$ as
$T\rightarrow T_2$, where $T_2$ is a second-order phase transition point.
For instance it has been found that the ratio
$T_g/T_2 = 1.28 \pm 5.8$\% in agreement with the AG prediction $T_g/T_K = 1.30 \pm 8.4$\%.
As may be seen from Table I, the static length of cooperativity
at $T_g$ obtained from Eq.\ (\ref{eq:2.12}) in fact is too small to define cooperative regions.
Adam-Gibbs arguments give rearranging units with at most 10 atoms near $T_g$ \cite{Sil99}.
Considerations based on mean-field models of structural glass transition, lead
to bigger cooperative regions, as described next.

\subsection{Discontinuous mean-field spin glass models}
\label{sec:2.2}
From the notion of droplet fluctuations together with the concept of random first-order
transition observed in discontinuous mean-field spin glass
models \cite{KTW89,KW87}, a correlation length is predicted that varies as $\xi_{\rm MF}\sim(T-T_K)^{-2/d}$,
where $d$ is the space dimensions. To illustrate this mean-field result we follow the more
direct approach presented in \cite{KW87,KT95}. An statistical mechanical analysis based
on replica method \cite{KW87,TK88} reveals in mean-field models of structural glass transition
(p-spin models, Potts glass, etc.)\ that there are in fact
two distinct transitions at $T_D$ and $T_K$ ($T_D>T_K$), the first of which at $T_D$ is
a dynamical transition akin to ideal mode coupling transition \cite{GS92}.
This dynamical transition at $T_D$ appears only in the mean-field,
and is reduced to a crossover temperature in finite dimensions
by the nucleation process or activated hopping between the multitude of the glassy
metastable states, $\mathcal{N}(T) \approx\exp[N \sigma(T)]$, in the range of temperatures $T_K<T<T_D$.
Thus, the configurational entropy, $S_c(T)\equiv k_B\ln\mathcal{N}(T)$, maintains its
extensive character for $T_K<T<T_D$, as $\mathcal{N}(T)$ is exponential in the
number of particles $N$ in this temperature range. However, as the thermodynamic transition
point $T_K$ is approached from above, the configurational entropy vanishes linearly,
$S_c=S_{\infty}[(T/T_K) - 1)]$, and the system stays in a glassy configuration indefinitely.
It should also be pointed out that the replica overlap order parameter changes
discontinuously despite the lack of a latent heat.
The replica approach \cite{KW87,TK88} further indicates that
\begin{eqnarray}
\mathcal{N}(T<T_K) & \propto & \exp\left(\frac{T_K}{T_K-T}\right), \nonumber \\
\mathcal{N}(T>T_K) & \propto & \exp\left[N\left(\frac{T}{T_K}-1\right)\right].
\label{eq:2.21}
\end{eqnarray}
A correlation length for the thermodynamic transition at $T_K$ is obtained by
matching Eqs.\ (\ref{eq:2.21}) in the critical region, giving $N\sim (T-T_K)^{-2}$,
and letting $N\sim\xi_{\rm MF}^d$. Thus,
\be
\xi_{\rm MF}(T)=\xi_{\infty}[1-(T_K/T)]^{-2/d}.
\label{eq:2.22}
\ee
The correlation length exponent predicted is $\nu=2/3$ in $d=3$ dimensions.

We note, however, that the cooperative regions obtained from
Eq.\ ({\ref{eq:2.22}) involve at most 90 particles
at $T_g$ \cite{XW00}. This is a significant improvement over the AG result,
but falls short of the size of dynamic heterogeneous domains observed at $T_g$,
which typically comprise a few hundred atoms \cite{Don01}.
Another difficulty with this mean-field picture is that its glassy behavior is
greatly modified in the short-range versions when studied numerically in three
dimensions \cite{CCP98,CC01}. MC simulations of a short-range version of p-spin glass
model in three dimensions \cite{CCP98}, indicate a continuous/second-order transition
to the glassy phase accompanied by a diverging susceptibility or correlation length.
The finite-size scaling results are found to be in favor of a correlation length
exponent $\nu=1$ (see,  Fig.\ 3 in \cite{CCP98}). Similar behavior is observed in a
short-range Frustrated Ising Lattice Gas (FILG) model \cite{CC01}, which also is known to have
a random first-order transition in the mean-field. The correlation length exponent
is found to be unity for FILG too. In view of Harris criterion
for the relevance of disorder in a second-order phase transition \cite{Har74},
which requires $\nu\geq 2/d$, and the above observations,
we believe that mean-field $\nu =2/3$ should be regarded as a lower-bound estimate
of the correlation length exponent for the fragile glass-forming liquids.
In section \ref{sec:3}, we shall also argue for a correlation
length exponent $\nu=1$ for the fragile systems, using a different argument.

\subsection{Dynamics as an activated potential energy barrier crossing}\label{sec:2.3}
It is a long held view that in supercooled liquids, dynamics is dominated
by the topographic properties of the system's Potential Energy Landscape (PEL) \cite{Gol69,SW82}:
long-time $\alpha$ relaxation is dictated by thermally activated crossing of the potential
energy barriers separating different valleys of the potential energy surface
$\Phi(\rb_1,\cdots,\rb_N)$, which is defined over the 3N-dimensional configurational
space of the liquid comprising $N$ atoms. It is also argued that activated transport
over the potential energy barriers begins to dominate at low temperatures, where
$E_{\rm eff}(T)\gtrsim 5k_BT$ \cite{Gol69}.
This description of dynamics in terms of the (3N+1)-dimensional
PEL, facilitates the study of collective phenomena in viscous liquids, and helps
to unify in a simple way some of the static and kinetic phenomena associated with the
glass transition \cite{Sti95}.

More recently, it has been demonstrated using molecular dynamics simulations of model
glass-forming liquids that the concept of activated hopping between whole super-structures
of many PEL minima, called PEL Metabasins (MBs), is central to a quantitative
description of the long-time dynamics in glass-forming liquids \cite{BH99,DH03}.
Here, the time evolution of the system is regarded as a sequence of MB visits each
with a residence time $\tau$. The mean residence/escape time from MB of energy $e$ is given by
$\langle\tau(e,T)\rangle = \tau_{\infty} \exp(E(e)/k_BT)$, where $e$ is defined as the
energy of the lowest local minimum within the MB. In fact the activation barrier, $E(e)$,
is found to depend on the depth of MB in PEL, $e$, in a rather simple way \cite{DH03,SDH03}:
$E(e) \approx -e$. The lower the $e$, the higher is the activation barrier $E(e)$.
A suitable average over the MBs visited by the representative point at a given temperature,
gives the average residence time $\langle\tau(T)\rangle$, which is to be regarded as
$\tau_{\alpha}(T)$:
\begin{eqnarray}
\tau_{\infty}/\langle\tau(T)\rangle &   =    & \int de\;p(e,T) \exp(-E(e)/k_BT) \nonumber \\
                                    & \equiv & \exp(-E_{\rm eff}(T)/k_BT),
\end{eqnarray}
where $p(e,T)\;de$ is the fraction of MBs within the range $e$ to $e$+$de$ visited by the
representative point at $T$. Clearly, $E_{\rm eff}(T)$ may be interpreted as some suitable
average over the potential energy barriers $E(e)$ ($\approx -e$) encountered by the liquid
at a given temperature \cite{DH03}. An spectacular demonstration of this assertion is the concurrence
between the crossover to super-Arrehnius relaxation and the commencement of the variation with
temperature of $\langle e(T)\rangle$ in an 80:20 BMLJ model liquid (see, Fig.\ 1 in \cite{Sas99}).
In order to better illustrate the close correlation between $E_{\rm eff}(T)$ and
$\langle e(T)\rangle$, in Fig.\ 1 we plot $k_BT\ln(\tau_{\alpha}/\tau_{\infty})$
($\equiv E_{\rm eff}(T)$) against the average value of PEL minima, $\langle e(T)\rangle$,
for the temperatures accessed using the data of Ref.\ \cite{Sas99}. The correlation is
rather impressive. This indicates among other things that the effective activation
energy barrier embodied in the empirical VF equation, may also be taken as an estimate for
the thermal variation of $\langle e(T)\rangle$, or that of the configurational energy density
$\langle\Phi(T)\rangle/N$, of the liquid \cite{Dav05}:
\be   E_{\rm eff}(T)\sim -\langle\Phi(T)\rangle/N.\label{eq:2.31}\ee
We use this result to estimate the temperature variation of the structural
specific heat for the fragile glass-forming liquids.

\section{Implication of Bond Ordering Scenario for static length of relaxation}
\label{sec:3}
In an attempt to clarify the physical origin of the dynamic heterogeneity observed
in deeply supercooled liquids, we expand on the concept of bond ordering \cite{DP01} and the
associated correlation length \cite{Dav05}. As pointed out in section \ref{sec:2},
$\xi_{\rm MF}(T_g)$ is a significant improvement over $\xi_{\rm AG}(T_g)$,
but consistently smaller than the typical linear size of the dynamic heterogeneous domains,
$\xi_{\rm het}(T_g)$. As it turns out another static correlation length, $\xi_{\rm BO}(T_g)$,
defined as the length-scale over which the potential energy fluctuation is correlated,
is in perfect agreement with the experimental observation.

By bond ordering we refer to the correlated relaxation of bonds into their low-lying energy
states, where inter-atomic bonds (as opposed to atoms) are treated as distinct objects possessing
internal degrees of freedom or energy levels. The length-scale over which the energy
fluctuations are correlated thus defines the {\em static length of relaxation}, $\xi_{\rm BO}$.
The {\em fourth-order} correlation function in terms of the local density $\rho (\rb)$,
which may also be interpreted as a two-point energy correlation function in terms of the
local potential energy density $\phi(\rb)$, is defined by \cite{Dav06}
\begin{eqnarray}
G_{\rm E} (\rb) & \equiv & \langle \phi(\rb) \phi(\Ob)\rangle - \langle \phi \rangle^2 \nonumber \\
                &  \sim  & g(r/\xi_{\rm BO})/r^{d-2+\eta'}.
\label{eq:3.1} \end{eqnarray}
$G_{\rm E} (\rb)$ is a fourth-order correlator as $\phi(\rb)\propto\rho(\rb)^2$.
The exponent $\eta'$, is related to the structural specific heat
exponent $\alpha$, via a fluctuation-response equation
\begin{eqnarray}
C_s & = & \frac{1}{k_BT^2}\int d^d r G_{\rm E}(\rb) \nonumber \\
    & \sim & \int^{\xi_{\rm BO}} d^d r/ r^{d-2+\eta'} \sim\xi_{\rm BO}^{2-\eta'},
\label{eq:3.2} \end{eqnarray}
where $\xi_{\rm BO}$ is the static correlation length beyond which the correlation
function rapidly vanishes. From Eq.\ (\ref{eq:3.2}), where the link
between the structural specific heat $C_s$ and the static relaxation length
$\xi_{\rm BO}$ becomes apparent, an energy version of the Fisher scaling
\cite{Fis64} is obtained: \be \alpha = (2-\eta')\nu. \label{eq:3.3} \ee
Assuming $\eta'=0$ for the fragile glass-forming liquids, which has been
corroborated by the numerical studies of short-range versions of mean-field
structural glass models \cite{CCP98,CC01}, we get
\be \nu = \alpha/2. \label{eq:3.4}\ee

The structural specific heat $C_s$ is in fact the temperature rate of
change of the configurational energy. In view of Eq.\ (\ref{eq:2.31}),
it is approximated by \be C_s = -\frac{\partial
E_{\rm eff}}{\partial T}\;. \label{eq:5.29} \ee
Using the effective energy barrier implied by
equation (\ref{eq:1.2}), i.e., $E_{\rm eff}(T) = Ak_B T/(T-T_0)$,
we have \be C_s = A k_B T_0/(T-T_0)^2. \label{eq:5.32}\ee
Eq.\ (\ref{eq:5.32}) implies a power law temperature variation
for the structural specific heat of the fragile supercooled
liquids, with an exponent $\alpha = 2$.
A critical power-law increase for the structural specific heat, has been recently
reported in MC simulations of a BMLJ model liquid \cite{Fer05}.
The exponent $\nu$ is thus given by $\nu = \alpha/2 = 1$.
Hence, with the effective potential energy barrier embodied in the
empirical Vogel-Fulcher equation, we have
\be \xi_{\rm BO}(T) = \xi_{\infty} T/(T-T_0). \label{eq:5.34} \ee
In fact, the quantity $F \equiv T_g/(T_g-T_0)$ ($=\xi_{\rm BO}(T_g)/\xi_{\infty}$)
is some times regarded as a
measure of the fragility of the liquid \cite{Ang91}. Its range of values for intermediate to
most fragile of the liquids is $3< F < 7.5$. Thus, more fragile the liquid,
the larger is the static length of relaxation at $T_g$.
Using Eq.\ (\ref{eq:5.34}), and $T_g/T_0 = 1.28\pm 5.8\%$, the static length of relaxation
at $T_g$ is determined to be in the range $3.4$--$7.0$ inter-atomic
spacings. This is indeed in perfect agreement with the typical linear size of the dynamic
heterogeneous domains observed at $T_g$ \cite{Edi00,Ber05}.

\section{Discussion}\label{sec:4}
In Table I we compare the values obtained for static lengths $\xi_{\rm BO}$,
$\xi_{\rm MF}$ and $\xi_{\rm AG}$ at $T_g$. This is done using the ratio
$T_g/T_0 = 1.28\pm 5.8\%$ \cite{Moh94,BC64}. We find $\xi_{\rm BO}(T_g)=3.4$--$7.0$
atomic diameters
in excellent agreement with $\xi_{\rm het}(T_g) \gtrsim 5 $ atomic diameters.
We also note that (energy) fluctuations of all linear sizes $x\lesssim\xi_{\rm BO}$
must be possible, and indeed probable.
(A similar, but stronger, effect arising from density fluctuations is believed to be
the cause of critical opalescence observed in fluids near their critical point \cite{Stan}.)
Thus, one expects heterogeneous domains with a distribution of relaxation times
at $T_g$, or not too far above $T_0$, such that the bigger domains have longer life-times
$\tau$. By assuming an activated form $\tau\sim e^x$, and $x$ to vary by a factor of 10,
we have $\tau$ values that vary by 4 decades. This therefore explains the existence of
dynamic heterogeneities near $T_g$ whose life-times differ
by several orders of magnitude \cite{Edi00}.
We see that the correlated relaxation of bonds in terms of energy, can be viewed as the
physical origin of the dynamic heterogeneity observed in deeply supercooled liquids.

It also is a matter of considerable interest that the BO exponent, $\nu=1$,
is in perfect agreement with that obtained from three-dimensional simulations of models
of structural glass that exhibit a random first-order transition in the mean-field \cite{CCP98,CC01}.
In fact, it appears that bond ordering scenario provides a physical
interpretation for the replica symmetry breaking transition observed in theory and simulation,
where members of coupled replica tend to sit in front of each other in the low temperature glassy phase.
Let us consider $m$ coupled replica constrained to be in the same state. This
resembles a liquid consisting of (super)molecular structures with every molecule
composed of $m$ number of coupled atoms \cite{Mez01}. This is an artifact of the attractive coupling among the
replica, and can be viewed as a theoretical indication for the local ordering of bonds in the realistic systems.
A detailed investigation of this apparent analogy will be presented elsewhere.

\section{Summary}\label{sec:5}
To summarize, the exponent governing the thermal behavior of the
static relaxation length in fragile supercooled liquids, $\nu=1$,
implies a stronger temperature dependence for cooperativity than
previously thought. The static relaxation length estimated at
$T_g$ is in perfect agreement with the typical size of the
dynamic heterogeneous domains in deeply supercooled liquids.
Furthermore, this larger static length also explains the wide
distribution of relaxation times associated with the
heterogeneities, which is central to explaining the stretched
exponential relaxation \cite{XW01}, and the decoupling of
self-diffusion from viscosity in deeply supercooled liquids
\cite{CCFN99}. The correlated relaxation of bonds in terms of
energy is therefore identified as the origin of dynamic
heterogeneity in fragile supercooled liquids.

\clearpage

\begin{figure}
\includegraphics[width=\textwidth]{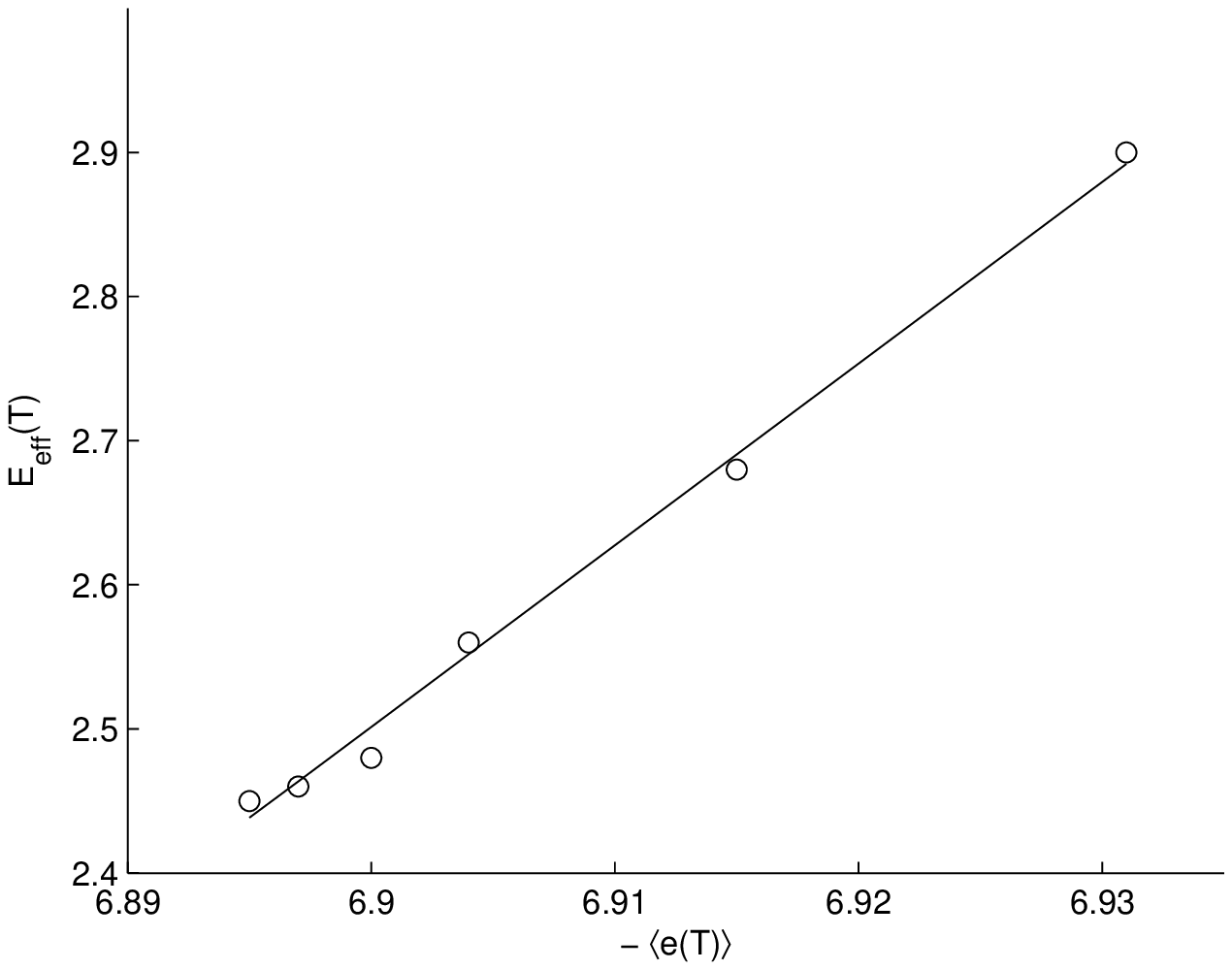} \caption{Dynamical activation energy 
$E_{\rm eff}(T)$ is plotted against
the average depth of PEL $\langle e(T)\rangle$ for temperatures
accessed using the data of Ref.\ \cite{Sas99} in order to
illustrate their correlation. Solid line is the least squares fit
to the data points.}
\end{figure}

\begin{table}
\caption{The static correlation lengths $\xi_{\rm BO}$, $\xi_{\rm
MF}$ and $\xi_{\rm AG}$ at $T_g$ are tabulated for comparison.
The observed linear size of dynamic heterogeneous domains
\cite{Edi00,Ber05}, $\xi_{\rm het}(T_g)/\xi_{\infty} \gtrsim 5$,
is in excellent agreement with the prediction based on bond
ordering picture.} \label{table1}
\begin{ruledtabular}
\begin{tabular}{lcc}
Model&   $\nu$&    $\xi(T_g)/\xi_{\infty}$ \\
\colrule
Bond Ordering&       1&  3.4--7.0       \\
Mean Field&       2/3&  2.3--3.7     \\
Adam and Gibbs&       1/3&  1.5--1.9   \\
\end{tabular}
\end{ruledtabular}
\end{table}

\end{document}